\documentclass[smallextended]{svjour3} 
\usepackage{graphicx}
\usepackage{amsmath}
\usepackage{float}
\usepackage{wrapfig}
\usepackage[justification=justified, font=footnotesize, width=0.8\textwidth]{caption}
\usepackage{tabularx}
\usepackage{relsize}
\usepackage{color}
\usepackage{bm}
\usepackage[normalem]{ulem}
\usepackage{cite}				% to join consecutive citations
\usepackage{tikz}	            % to draw on tables
\definecolor{carmin}{rgb}{0.4,0,0.2}
\newcommand\diag[4]{%
\multicolumn{1}{p{#2}|}{\hskip-\tabcolsep
$\vcenter{%
\begin{tikzpicture}[baseline=0,anchor=south west,inner sep=#1+5pt]
  \path[use as bounding box] (-0.007,36.25pt) rectangle (#2+2\tabcolsep,\baselineskip);
  \node[minimum width={#2+2\tabcolsep-\pgflinewidth},
        minimum  height=\baselineskip+20pt-\pgflinewidth+6.93pt] (box) {};
  \draw[line cap=round] (box.north west) -- (box.south west);
  \draw[line cap=round] (box.north west) -- (box.south east);
  \node[anchor=south west] at (box.south west) {#3};
  \node[anchor=north east] at (box.north east) {#4};
\end{tikzpicture}}$\hskip-\tabcolsep}}

\newcommand\cen[4]{%
\multicolumn{1}{p{#2}|}{\hskip-\tabcolsep
$\vcenter{%
\begin{tikzpicture}[baseline=0,anchor=south west,inner sep=#1]
  \path[use as bounding box] (-0.007,36.25pt) rectangle (#2+2\tabcolsep,\baselineskip);
  \node[minimum width={#2+2\tabcolsep-\pgflinewidth},
        minimum  height=\baselineskip+20pt-\pgflinewidth+0.35pt] (box) {};
  \draw[color=gray, dashed] (0.6,0.05) -- (0.6,1.29) ;
  \node[anchor=west] at (box.west) {#3};
  \node[anchor=east] at (box.east) {#4};
\end{tikzpicture}}$\hskip-\tabcolsep}}

\begin{document}

\title{Quantum Locality, Rings a Bell?:\\ 
Bell's inequality meets local reality and true determinism}

\author{S\'anchez-Kuntz, Natalia  \and   Nahmad-Achar, Eduardo}

\institute{Nahmad-Achar, Eduardo \at
Instituto de Ciencias Nucleares, Universidad Nacional Aut\'onoma de M\'exico \\
Apdo. Postal 70-543 M\'exico, Cd.Mx., 04510 \\
	              Tel.: +52 (55) 56224660 ext 2263\\
	              Fax:  +52 (55) 56224682\\
	              \email{nahmad@nucleares.unam.mx}           
	           \and
	          S\'anchez-Kuntz, Natalia \at
				 Instituto de Ciencias Nucleares, Universidad Nacional Aut\'onoma de M\'exico \\
				 Apdo. Postal 70-543 M\'exico, Cd.Mx., 04510}

\date{Received: date / Accepted: date}

\maketitle

\section*{Acknowledgements}
\noindent
This work was partially supported by Direcci\'on General de Asuntos del Personal Acad\'emico, Universidad Nacional Aut\'onoma de M\'exico (under Project No.~IN101217). NS-K thanks Consejo Nacional de Ciencia y Tecnolog\'ia-M\'exico for financial support.

\newpage
\begin{abstract}
	By assuming a deterministic evolution of quantum systems and taking realism into account, we carefully build a hidden variable theory for Quantum Mechanics based on the notion of ontological states proposed by \,'t Hooft\cite{tHooft}. We view these ontological states as the ones embedded with realism and compare them to the (usual) quantum states that represent superpositions, viewing the latter as mere information of the system they describe. 

	Such a deterministic model puts forward conditions for the applicability of Bell's inequality: the usual inequality cannot be applied to the usual experiments. We build a Bell-like inequality that can be applied to the EPR scenario and show that this inequality is always satisfied by Quantum Mechanics. 

	In this way we show that Quantum Mechanics can indeed have a local interpretation, and thus meet with the causal structure imposed by the Theory of Special Relativity in a satisfying way.

\keywords{Foundations of quantum mechanics \and Quantum locality \and Hidden variables}
\PACS{03.65.Ta \and 03.65.Ud \and 03.65.Ca}
\end{abstract}

\section{Introduction}

\noindent Since Einstein, Podolsky, and Rosen questioned the nature of Quantum Mechanics and its predictions \cite{EPR}, the quest for an interpretation of the paradoxical aspects they pointed out has taken a wide variety of paths \cite{tHooft,Peres,PBR,Lundeen,Spekkens}. No agreement has been reached, however. Is a quantum state real, or is it a carrier of information? Is the wave function only a mathematical construct, even when we can see wave-like interference patterns in Young's double slit experiment? Quantum states in superposition cannot be observed (the dead-and-alive cat, for instance) suggesting that they merely embody statistical restrictions on measurement results. Yet we think of them as describing physical systems that evolve in time in accordance to well given mathematical equations. 

This evolution, we picture, takes place in physical spacetime, and this spacetime is endowed with a locally causal structure.
But there is a violation of causality embedded in Quantum Mechanics; so much so that many interpretations have been given as to what this violation might physically mean\cite{Maudlin, Aharonov, Leifer, Cabello}. Local causality is imposed on spacetime by Special Relativity: a sequence of cause and effect that constitutes, we believe, a fundamental principle on which we think about and do our scientific work.

This means that we need a better understanding of the most basic phenomena of Quantum Mechanics. Several no-go theorems have shut the door for realism and locality\cite{Bell,Peres,GHZ,KS,CHSH}; but in which way?, with what assumptions?, is the door really locked? In this work we will start to examine these questions by proposing a realist hidden variable interpretation of Quantum Mechanics: \emph{factuality}. Within this perspective we will analyse the first and most important of the no-go theorems: Bell's inequality\footnote{Although the inequality that is experimentally tested\cite{Aspect, Aspect1} is the variation of Bell's inequality formulated by Clauser \emph{et. al.}\cite{CHSH}, we will revisit Bell's original construction\cite{Bell}, given that the analysis we make rests on the common ground of both, and it is easier to look at the original one.}. This is only a first step towards developing a local deterministic formulation of Quantum Mechanics. 

\section{Construction}

We will begin by revisiting the tools of Quantum Mechanics that are necessary for the construction of our proposal. To do so, we will make a general statement that will be applied to the particular case of a spin degree of freedom for fermions (which might be extended to polarisation for photons).

\subsection{Tools and ontology}

Quantum Mechanics (QM) is a wave theory in that it associates wave properties to particles. But it actually reduces all mechanics to the mechanics of particles themselves. The wave nature (as in the double-slit experiment) arises when one observes the statistical behaviour of a large ensemble of particles, just as ripples in water arise from a statistical behaviour of many water particles, or electromagnetic waves, in quantum theory, are the result of a large collection of photons. We see the phenomenon of superposition in waves, but not in the individual particles which are the building blocks (physical entities) in QM. With this in mind, we can then say that:

\begin{quote}
Physical entities do not appear in superposed states, that is, \emph{nature in its fundamental parts does not emerge as a superposition of states}. The superposition principle is a mathematical construct which can then be applied to the individual parts of an ensemble as a statistical description of the ensemble, not to each individual entity as a realistic description of the latter.
\end{quote}

We will also recall the way Bohr\cite{Bohr} interpreted the uncertainty principle: he ascertained that non-commuting operators defined realities that would appear in a complementary manner, that is, each one in its own and excluding frame of reference. Following this notion, we regard a frame of reference as that one which is determined by a complete set of commuting operators. 

With these two statements in hand, we define:

%                                                  %%%
\begin{quote}
\textbf{Quantum states} as the states generated from linear combinations of different eigenstates of an observable (not all of them with the same eigenvalue), and denote them by $\vert \psi \rangle$. 

\textbf{Ontological states} as the eigenstates of a complete set of mutually commuting observables, and denote them by $\vert \Omega \rangle$.
\end{quote}
%                                                  %%%
So, in any frame of reference, ontological states would be those accessible to the system, the \emph{real states} that the system could be in, while quantum states would be the only available ones we have to describe reality, due to a lack of knowledge of the complete and deterministic evolution of any state.

It is important to notice that, if we have a quantum state description of a system, we can always perform a basis transformation so that this description becomes an ontological state description. For example: $\frac{1}{\sqrt{2}}\left[\hspace{1pt} \vert \hspace{-1pt} \uparrow  \hspace{1pt} \rangle^{z}+\vert \hspace{-1pt} \downarrow  \hspace{1pt} \rangle^{z} \right]$ is a quantum state description of the observable
$\hat{\sigma}_z$, but acquires an ontological meaning when we switch to the $\hat{\sigma}_x$-diagonal basis\footnote{That is, $\frac{1}{\sqrt{2}}\left[ \hspace{1pt} \vert \hspace{-1pt}  \uparrow  \hspace{1pt} \rangle^{z}+\vert \hspace{-1pt} \downarrow \hspace{1pt} \rangle^{z} \right]$ is ontological when the chosen set of commuting observables is $\lbrace S^2 , S_x \rbrace$, rather than $\lbrace S^2 , S_z \rbrace$.}, resulting in $\vert  \hspace{-1pt}  \uparrow  \hspace{1pt} \rangle^{x}$.

On the other hand, ensembles of individual particles might be described either as pure states or as mixed states. Of course, each description depicts different ensembles. In a pure state description we regard the ensemble as if every one of its components were in precisely that pure state, while in a mixed state description we regard the ensemble as one where different components of the ensemble are in different pure states, with a certain probability distribution. We will denote these two descriptions as $\rho$ and $\tilde{\rho}$ respectively, i.e.,
%                                                  %%%
$$
\rho=\vert \Omega \rangle \langle \Omega \vert\,, 
$$

\vspace{-18pt}

$$
\tilde{\rho}=\sum_{i} c_i\vert \Omega_i\rangle \langle \Omega_i \vert\,.
$$
%                                                  %%% 
Along the same line in which quantum states emerge only as a mathematical description of a system, mixed states only represent a statistical description of an ensemble that is comprised of many entities, each one in a pure state. 

\textbf{Ontological pairs} are those which emerge due to the interaction between $A$ and $B$, two physical entities (be them $A$, an electron and $B$, a measurement device; or $A$ and $B$ two electrons in a spin state $S=0$; or any two particles $A$ and $B$ that come together at time $t=t_0$). 

An ontological pair is the complete and known description of a system at a given time, $t_0$. For example, in a measurement of any given property, what we describe (and know) is the ontological pair of the system [particle]-[measurement device].

Ontological pairs can exhibit entanglement. Entangled states give rise to non-separable pure states. These will be denoted by:
$$
{}_{\tilde{A}}\rho_{\tilde{B}}= \Big[\vert \phi \rangle \langle \phi \vert \Big]_{\tilde{AB}}\,.
$$
%                                                  %%%

As we will see in the following section, we propose that ontological states evolve according to a function of time and a hidden variable.

\subsection{Evolution}

In our construction, beyond the realism embedded in the ontological state description that we put forward above, we must have a deterministic evolution of the ontological pairs, and thus we should give a function that governs such evolution. 

We are analysing the thought experiment of Bohm and Aharonov\cite{Bohm}, so we work with an ensemble of two-body systems in an entangled state
%                                                  %%%
$$
\vert \phi \rangle_{AB} = \frac{1}{\sqrt{2}} \left[ \vert \hspace{-2pt} \uparrow \downarrow \rangle^{\vec{z}} +  \vert \hspace{-2pt} \downarrow \uparrow \rangle^{\vec{z}} \right]\,.
$$
%                                                  %%%
Each system divides into its two components and each of these reaches a detector, where its spin projection is measured. 

In our view each individual system is an ontological pair, which will evolve according to a function of a hidden variable $\lambda$ and time $t$. In this particular case we are only focusing on the projection of the spin degree of freedom of each component of the pair, then such function, when evaluated at a given value of the hidden variable $\lambda$ and the time of measurement $t_1$, will result in the direction of the spin projection of the two components of the ontological pair. That is: 
%                                                  %%%
$$
\mathcal{F} : \Lambda \times \mathbf{R} \longrightarrow \mathbf{R^3} \times \mathbf{R^3}\,,
$$
%                                                  %%%
where $\Lambda$ is the set of values that the hidden variable can take, i.e., it is the domain of $\lambda$. So, given $\lambda \in \Lambda$ and the time of measurement, $t_1 \in \mathbf{R}$,
%                                                  %%%
$$
\mathcal{F} (\lambda,t_1)=(\vec{o}_{_A},\vec{o}_{_B})\,,
$$
%                                                  %%%
where $\vec{o}_{_A}$ and $\vec{o}_{_B}$ are the spin projection orientations of each component of the pair, and they themselves are functions of $\lambda$ and $t_1$, $\vec{o}_{_A}(\lambda,t_1)$ and $\vec{o}_{_B}(\lambda,t_1)$. Note that these functions are absolutely deterministic, and a direct consequence of this is the fact that the orientation of the detectors is also encoded in $\lambda$. There is no \emph{what would have happened if the detector had not been in such and such orientation?} The detector will have only one true orientation, determined by all the previous conditions accessible to it. This is what a truly deterministic scenario entails. A detector in a different orientation will have different values of $\lambda$ at all earlier times.

Remaining within our description, the spin degree of freedom of a two-body system would evolve from one ontological state to the next, while there is a change in frame of reference. The initial ontological state and frame of reference, being:
%                                                  %%%
$$
\vert \Omega(t_0)\rangle = \vert \phi \rangle_{AB} = \frac{1}{\sqrt{2}} \left[ \vert \hspace{-2pt} \uparrow \downarrow \rangle^{\vec{z}} +  \vert \hspace{-2pt} \downarrow \uparrow \rangle^{\vec{z}} \right]
$$
%                                                  %%%
and
%                                                  %%%
$$
\lbrace (S_A+S_B)^2 , (S_A+S_B)_{\vec{z}} \rbrace\,,
$$
%                                                  %%%
and the final ontological state and frame of reference being:
%                                                  %%%
$$
\vert \Omega(t_1)\rangle = \vert \uparrow\, \uparrow \,\rangle^{\mathcal{F} (\lambda, t_1)}
$$
%                                                  %%%
and
%                                                  %%%
$$
\lbrace (S_A)^2, (S_B)^2 , (S_A)_{\vec{a}}, (S_B)_{\vec{b}} \rbrace\,,
$$
%                                                  %%%
where $\vec{a}$ and $\vec{b}$ are the two orientations of the detectors over particle $A$ and $B$, respectively. 

These are all the tools we need for an ontological and deterministic description of reality. In the next section we will talk about locality conditions and the mechanism for entanglement. 

\section{Locality}
We have constructed a description of entanglement that is implicitly local, given the introduction of hidden variables. We make, though, one statement about deterministic evolution that was not made when hidden variable interpretations were first introduced\cite{Bell} and then we put forward the mechanism for entanglement.   

We affirm that the evolution function $\mathcal{F}(\lambda,t)$ must satisfy a condition we call {\bf factuality}. Mathematically, this condition is no news: for any given function, different outcomes of the function must come from different inputs. So, once the values of hidden variable and time are given, our function $\mathcal{F}(\lambda, t)$ can only acquire a certain value $(\vec{o}_A,\vec{o}_B)$.
Physically, this is the factuality condition: if a system evolved in time ($t_0 \rightarrow t_1$) to a particular state, it is because only this state was accessible to it given the initial condition $(\lambda,t_0)$ and, therefore, different states at time $t_1$ must come from different values of hidden variables $\lambda_i$. This is only a consequence of determinism.

In our view, non-local correlations emerge from the deterministic evolution of a shared hidden variable between two components of an ontological pair. Entanglement arises every time two (or more) physical entities share hidden variables. This suffices for the time being, and for the example we work below. In what follows, we will analyse the emergence of Bell's inequality within our proposed description of reality. 

\section{Bell's inequality}

\noindent Suppose a pair of entangled electrons in a singlet state is split into two electrons at time $t=t_0$. If the spin of electron $A$ is measured at a later time in the $\vec{z}$ direction and we get, for example, $\vert \hspace{-3pt} \uparrow \rangle_A^{\vec{z}}$ then we can be sure that the spin of electron $B$ is  $\vert \hspace{-2pt} \downarrow \rangle_B^{\vec{z}}$. Locality associates, with the spin of each electron, a hidden variable quality, that is: $A(\vec{a}, \lambda)=\pm 1$ where $A$, the value of the spin of particle $A$, is a function of the direction of the detector $\vec{a}$ and of a hidden variable $\lambda$. Same for $B$ in any direction $\vec{b}$. The expectation value of the correlation between $A$ (measured in the direction $\vec{a}$) and $B$ (measured in the direction $\vec{b}$) naturally arises,
%                                                  %%%
$$
E(\vec{a}, \vec{b})= \int_{\Lambda} A(\vec{a}, \lambda) B(\vec{b}, \lambda) \rho({\lambda}) d\lambda\,.
$$
%                                                  %%%
Bell shows\cite{Bell} that if such functions $A$ and $B$ exist, the expectation value of the correlation between them must satisfy: 
%                                                  %%%
$$
\vert E(\vec{a}, \vec{b})-E(\vec{a}, \vec{c}) \vert \leq 1 + E(\vec{b}, \vec{c})\,,
$$
%                                                  %%%
where $\vec{a}$, $\vec{b}$, and $\vec{c}$ are three alternative directions of the detectors used to measure the spin projection of the electrons. 

\subsection{Under factuality}
First we need to find the relation between the functions $A(\vec{j}, \lambda)$ and $B(\vec{k}, \lambda)$ and our deterministic evolution function $\mathcal{F} (\lambda,t)$.

Functions $A$ and $B$ are both results of a measurement, so they must be related to the function $\mathcal{F}$ when the latter is evaluated at the time of measurement, $t=t_1$.

Now, $\mathcal{F}(\lambda,t_1)$ gives a pair of orientations, $(\vec{o}_A,\vec{o}_B)$. These two orientations are those of the spin projection for particles $A$ and $B$ respectively at the time of measurement, and it is important to recall that the orientation of the two detectors is also encoded in the value of the hidden variable $\lambda$. 

Function $A(\vec{j}, \lambda)$ asks the question, ``given a detector device with orientation $\vec{j}$ and a hidden variable $\lambda$, is the electron's spin orientation $\vec{j}$ or $-\vec{j}$?''. So for this question to be posed, the electron's spin orientation \emph{must} be $\vec{j}$ or $-\vec{j}$. Analogously for function $B(\vec{k}, \lambda)$. Then, these two questions can be posed iff $\mathcal{F} (\lambda,t_1)=(\pm\vec{j},\pm\vec{k})$. 

\begin{quote}
\textbf{Fact 1} functions $A(\vec{j}, \lambda)$ and $B(\vec{k}, \lambda)$ are simultaneously well defined iff $\mathcal{F}(\lambda,t_1)=(\pm\vec{j},\pm\vec{k})$.
\end{quote}

%                                                  %%%            figure
\begin{figure}[htb]
\begin{center}
	\includegraphics[width=0.8\textwidth]{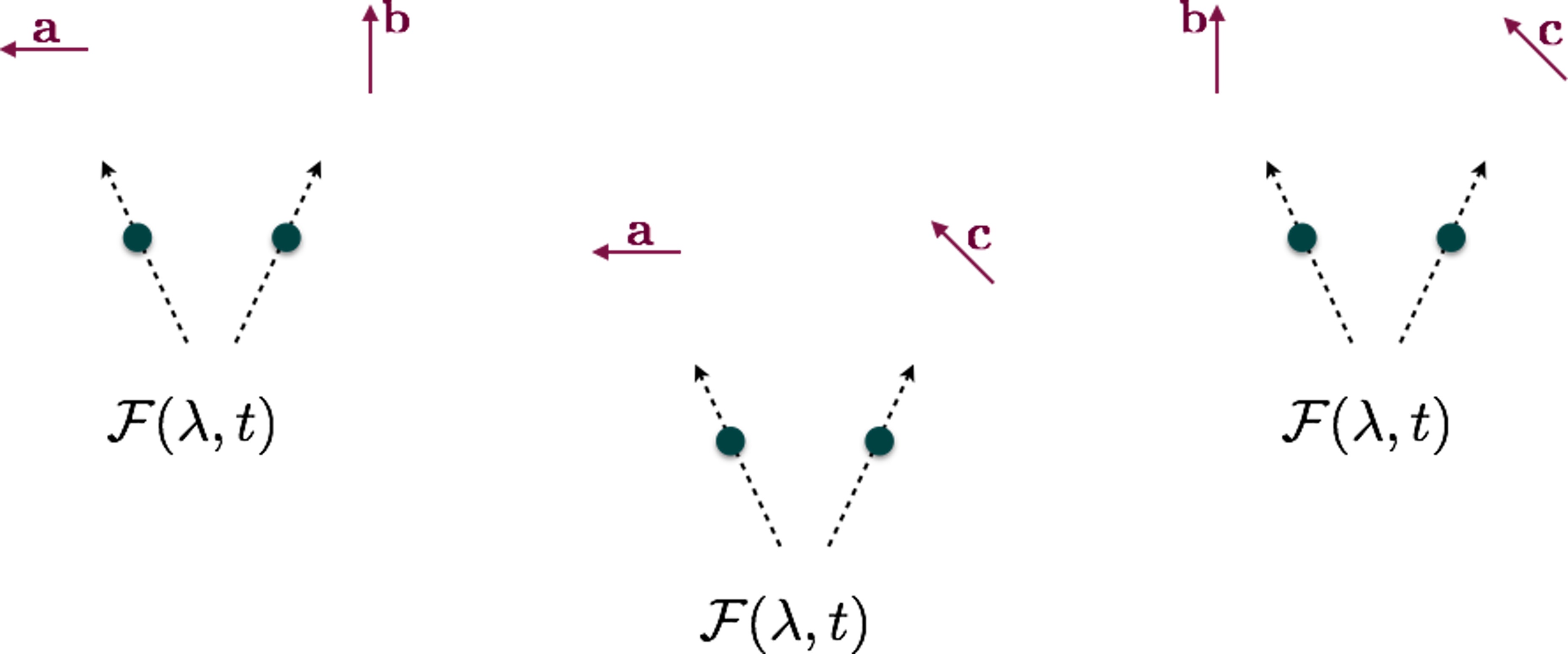}	
	\caption{Three ontological pairs whose spin degree of freedom is described by the function $\mathcal{F}(\lambda,t)$, each pair subject to a different set of measurements at time $t=t_1$.} 
	\label{exp}
\end{center}
\end{figure}
%                                                  %%% 

Now we will impose the factuality condition on three deterministic scenarios (Fig. \ref{exp}). In the left hand side scenario of that figure, the measurement outcome can be any of four different possibilities, $(\vec{a},\vec{b})$, $(\vec{a},-\vec{b})$, $(-\vec{a},\vec{b})$ and $(-\vec{a},-\vec{b})$, that is $\mathcal{F}(\lambda,t_1)=(\pm\vec{a},\pm\vec{b})$. In the second scenario, $\mathcal{F}(\lambda,t_1)=(\pm\vec{a},\pm\vec{c})$ and in the right hand scenario $\mathcal{F}(\lambda,t_1)=(\pm\vec{b},\pm\vec{c})$.

Under the factuality condition, each of these sets of outcomes must come from a different set of hidden variables, that is:
%                                                  %%%            
\begin{equation}
\mathcal{F}(\lambda,t_1)=(\pm\vec{a},\pm\vec{b}) \, \leftrightarrow \, \lambda \in \Lambda_1\,,
\label{l1}
\end{equation}
%                                                  %%% 
\begin{equation}
\mathcal{F}(\lambda,t_1)=(\pm\vec{a},\pm\vec{c}) \, \leftrightarrow \, \lambda \in \Lambda_2\,,
\label{l2}
\end{equation}
%                                                  %%% 
\begin{equation}
\mathcal{F} (\lambda,t_1)=(\pm\vec{b},\pm\vec{c}) \, \leftrightarrow \, \lambda \in \Lambda_3\,.
\label{l3}
\end{equation}
%                                                  %%% 

Furthermore,  $\Lambda_1 \cap  \Lambda_2 =  \Lambda_1 \cap  \Lambda_3 =  \Lambda_2 \cap  \Lambda_3 = \emptyset$, which can be seen by the simple reasoning:

%                                                  %%%
\begin{center}
	If $\lambda \in \Lambda_1$,\hspace{2pt} then $\mathcal{F}(\lambda,t_1)=(\pm\vec{a},\pm\vec{b})\not=(\pm\vec{a},\pm\vec{c})$,\hspace{2pt} then $\lambda \notin \Lambda_2$; etc.
\end{center}
%                                                  %%%

Then, from \textbf{Fact 1} and equations (\ref{l1}), (\ref{l2}) and (\ref{l3}):

\begin{quote}
\textbf{Fact 2.1} functions $A(\vec{a}, \lambda)$ and $B(\vec{b}, \lambda)$ are simultaneously well defined iff $\lambda \in \Lambda_1$.
\end{quote}

\begin{quote}
\textbf{Fact 2.2} functions $A(\vec{a}, \lambda)$ and $B(\vec{c}, \lambda)$ are simultaneously well defined iff $\lambda \in \Lambda_2$.
\end{quote}

\begin{quote}
\textbf{Fact 2.3} functions $A(\vec{b}, \lambda)$ and $B(\vec{c}, \lambda)$ are simultaneously well defined iff $\lambda \in \Lambda_3$.
\end{quote}

So, if we were to follow Bell's steps to derive his inequality, we would start by comparing the expectation values,
%                                                  %%%
$$
E(\vec{a}, \vec{b})-E(\vec{a}, \vec{c}) = \int_{\Lambda_1} A(\vec{a}, \lambda) B(\vec{b}, \lambda)  \rho({\lambda}) d\lambda - \int_{\Lambda_2} A(\vec{a}, \lambda) B(\vec{c}, \lambda) \rho({\lambda}) d\lambda\,,
$$
%                                                  %%%
where we have explicitly written the integration domains imposed by \textbf{Fact 2.1} and \textbf{Fact 2.2}. 
Since $\Lambda_1 \cap \Lambda_2 = \emptyset$ we cannot carry on to Bell's next step in order to derive his inequality, so:

\begin{quote}
\textbf{Fact 3} In a local deterministic scenario, governed by factuality, Bell's inequality cannot be derived, therefore the violation of his inequality by experiments does not show that the assumption of locality in this scenario is incorrect.
\end{quote}

The statement above begs the question, in which scenario can Bell's inequality be derived? And, do experiments violate this inequality in such scenario? We will take a look at these questions in the next subsection.

\subsection{Building Bell's inequality}
\label{Blike}

\noindent In the previous sections we have worked with a deterministic view of reality, in which the spin degree of freedom of an ontological pair is governed by a function $\mathcal{F} (\lambda,t)$. 

In order to build Bell's inequality it is required that the set of hidden variables that lie behind the three different scenarios in Figure \ref{exp} be one and the same ($\lambda \in \Lambda$). If we want this requirement to be satisfied, we can take on two possible paths:

%This view can take on two possible paths when describing three different scenarios (cf. Figure \ref{exp}), they are: 

\begin{itemize}
\item Each different scenario can be governed by a different function, $\mathcal{F} _i(\lambda, t)$, $i = 1, 2, 3$.
\item On each different scenario the measurement can take place at a different time, so the final state could be described by $\mathcal{F} (\lambda, t_i)$, $i = 1, 2, 3$.
\end{itemize}

When taking any of these two paths, Bell's steps can be followed further than we could on the last subsection. As before, we will start by identifying the functions $A$ and $B$ used by Bell with our function $\mathcal{F} $.

We can directly see that functions $A(\vec{a},\lambda)$ and $B(\vec{b},\lambda)$ can only be simultaneously identified with $\mathcal{F} _1(\lambda, t_1)$ (in the first path) or $\mathcal{F} (\lambda, t_1)$ (in the second path). We will take on the first path (the second path is shown in Appendix A). 

We know that
%                                                  %%%            
$$
\mathcal{F} _1(\lambda, t_1) = (\vec{o}_{A_1}(\lambda, t_1),\vec{o}_{B_1}(\lambda, t_1)) = (\pm\vec{a},\pm\vec{b})\,,
$$
%                                                  %%%
then we can define
%                                                  %%%            
$$
A_1(\vec{a},\lambda) \equiv \rm{sign}(\vec{o}_{A_1}(\lambda, t_1))
$$
%                                                  %%%              
and 
%                                                  %%%            
$$
B_1(\vec{b},\lambda) \equiv \rm{sign}(\vec{o}_{B_1}(\lambda, t_1))\,.
$$
%                                                  %%%
Note that we carried the subscript $1$ to distinguish these functions from the ones defined by  $\mathcal{F} _2(\lambda, t_1)$. In this second case we have:
%                                                  %%%            
$$
\mathcal{F} _2(\lambda, t_1) = (\vec{o}_{A_2}(\lambda, t_1),\vec{o}_{B_2}(\lambda, t_1)) = (\pm\vec{a},\pm\vec{c})\,,
$$
%                                                  %%%
and we can simultaneously define
%                                                  %%%            
$$
A_2(\vec{a},\lambda) \equiv \rm{sign}(\vec{o}_{A_2}(\lambda, t_1))
$$
%                                                  %%%              
and 
%                                                  %%%            
$$
B_2(\vec{c},\lambda) \equiv \rm{sign}(\vec{o}_{B_2}(\lambda, t_1))\,.
$$
%                                                  %%%
And in the third case:
%                                                  %%%            
$$
\mathcal{F} _3(\lambda, t_1) = (\vec{o}_{A_3}(\lambda, t_1),\vec{o}_{B_3}(\lambda, t_1)) = (\pm\vec{b},\pm\vec{c})\,,
$$
%                                                  %%%
so
%                                                  %%%            
$$
A_3(\vec{b},\lambda) \equiv \rm{sign}(\vec{o}_{A_3}(\lambda, t_1))
$$
%                                                  %%%              
and 
%                                                  %%%            
$$
B_3(\vec{c},\lambda) \equiv \rm{sign}(\vec{o}_{B_3}(\lambda, t_1))\,.
$$
%                                                  %%%

\vspace{2pt}

Now that each scenario is governed by a different function  $\mathcal{F}_i$ we can go back to Bell's first step,

%                                                  %%%
$$
E(\vec{a}, \vec{b})-E(\vec{a}, \vec{c}) = \int_{\Lambda} A_1(\vec{a}, \lambda) B_1(\vec{b}, \lambda) \rho({\lambda}) d\lambda - \int_{\Lambda} A_2(\vec{a}, \lambda) B_2(\vec{c}, \lambda) \rho({\lambda}) d\lambda\,,
$$
%                                                  %%%

\vspace{5pt}

\noindent where we have implicitly written the subscripts that define each function $A_i$, $B_i$ in terms of the deterministic evolution function of each different experiment. 

And to his second step,
%                                                  %%%
\begin{equation}
\hspace{-0.7pt}\left| E(\vec{a}, \vec{b})-E(\vec{a}, \vec{c})\right|  = 
\left|  \int_{\Lambda}  [\bm{A_1(\vec{a}, \lambda)} B_1(\vec{b}, \lambda) -  \bm{A_2(\vec{a}, \lambda)} B_2(\vec{c}, \lambda)] \rho({\lambda}) d\lambda \right|,
\label{1}
\end{equation}
%                                                  %%%
where we have highlighted $A_1(\vec{a}, \lambda)$ and $A_2(\vec{a}, \lambda)$ to stress the fact that for his third step, Bell takes these two functions to be identical. This is his first assumption (out of three). We will analyze what can be said about the quantity $\left| E(\vec{a}, \vec{b})-E(\vec{a}, \vec{c})\right|$ in two cases: while taking Bell's three assumptions, and while taking none of them.

\vspace{5pt}
\underline{Within Bell's assumptions}

\vspace{10pt}
Bell's three assumptions are (shown in Appendix B):
%                                                  %%%
$$
A_1(\vec{a}, \lambda) = A_2(\vec{a}, \lambda)\,,
$$
%                                                  %%%
$$
B_1(\vec{b}, \lambda)=-A_3(\vec{b}, \lambda)\,,
$$
%                                                  %%%
$$
B_2(\vec{c}, \lambda)=B_3(\vec{c}, \lambda)\,.
$$
%                                                  %%%
These are constraints on the functions $\mathcal{F} _i(\lambda, t)$ that have to be met in order for Bell's  inequality to be derived. So, the applicable domain of his inequality is the one that behaves according to these constraints, that is, the deterministic functions $\mathcal{F} _i(\lambda, t)$ that govern the three experiments built to test Bell's inequality have to be so that these constraints are satisfied. 

This has an implication on the expectation values of the correlation between measurements. If these three constraints are satisfied, the predicted expectation values result in: 
%                                                  %%%
$$
E(\vec{a}, \vec{b})=-\cos\theta_{ab}\,,
$$
%                                                  %%%
$$
E(\vec{a}, \vec{c})=-\cos\theta_{ac}\,,
$$
%                                                  %%%
$$
E(\vec{b}, \vec{c})=-\cos\theta_{ab}\cos\theta_{ac}\,,
$$
%                                                  %%%
which is caused by the fact that the given constraints tamper with the probabilities of getting $(\pm \vec{a})$, $(\pm \vec{b})$ or $(\pm \vec{c})$ in the measurements performed. The derivation of these results is given in Appendix C.

Now, this result leads to two conclusions: 

The first one is: if the experiments were to satisfy the constraints necessary to build Bell's inequality, then the expectation values would be such that when plugged into the inequality one would get:
%                                                  %%%
\begin{equation}
\left|- \cos\theta_{ab}+\cos\theta_{ac}\right| \leq 1 - \cos\theta_{ab}\cos\theta_{ac}
\label{eq}
\end{equation}
%                                                  %%%
and, as shown in Appendix D, this inequality is always satisfied.

The second conclusion is: \textbf{the experiments} used to test Bell's inequality do not result in an expectation value given by a product of cosines\linebreak \noindent($-\cos\theta_{ab}\cos\theta_{ac}$), so they \textbf{do not behave according to the constraints necessary to build Bell's inequality}, so they do not have to satisfy such an inequality and the violation of the inequality by the experiments does not show that reality cannot behave in a local deterministic way.

\vspace{5pt}
\underline{Without Bell's assumptions}

\vspace{10pt}
We will now go back to his second step and build a Bell-like inequality without his assumptions. 
%                                                  %%%
$$
\left| E(\vec{a}, \vec{b})-E(\vec{a}, \vec{c})\right| = \left|  \int_{\Lambda}  [A_1(\vec{a}, \lambda) B_1(\vec{b}, \lambda) -  A_2(\vec{a}, \lambda) B_2(\vec{c}, \lambda)] \rho({\lambda}) d\lambda \right| 
$$
%                                                  %%%
$$   
\hspace{71pt}=\left| \sum_{i=1}^{8}  \int_{\tilde{\Lambda}_i}  [A_1(\vec{a}, \lambda) B_1(\vec{b}, \lambda) -  A_2(\vec{a}, \lambda) B_2(\vec{c}, \lambda)] \rho({\lambda}) d\lambda \right|,
$$
%                                                  %%%
where we build the sets $\tilde{\Lambda}_i$, $i=1,\ldots,8$, in terms of the different relations that the functions $A_1$, $A_2$, $A_3$, $B_1$, $B_2$ and $B_3$ hold between them. These 8 sets $\tilde{\Lambda}_i$ are defined as:

%                                                  %%%
$$
\tilde{\Lambda}_1=\{\lambda\, \vert \, A_1=A_2 \,,\, B_1=-A_3 \,,\, B_2=B_3\}\,,
$$
%                                                  %%%
$$
\tilde{\Lambda}_2=\{\lambda\, \vert \, A_1=A_2 \,,\, B_1=-A_3 \,,\, B_2=-B_3\}\,,
$$
%                                                  %%%
$$
\tilde{\Lambda}_3=\{\lambda\, \vert \, A_1=A_2 \,,\, B_1=A_3 \,,\, B_2=B_3\}\,,
$$
%                                                  %%%
$$
\tilde{\Lambda}_4=\{\lambda\, \vert \, A_1=A_2 \,,\, B_1=A_3 \,,\, B_2=-B_3\}\,,
$$
%                                                  %%%
$$
\tilde{\Lambda}_5=\{\lambda\, \vert \, A_1=-A_2 \,,\, B_1=-A_3 \,,\, B_2=B_3\}\,,
$$
%                                                  %%%
$$
\tilde{\Lambda}_6=\{\lambda\, \vert \, A_1=-A_2 \,,\, B_1=-A_3 \,,\, B_2=-B_3\}\,,
$$
%                                                  %%%
$$
\tilde{\Lambda}_7=\{\lambda\, \vert \, A_1=-A_2 \,,\, B_1=A_3 \,,\, B_2=B_3\}\,,
$$
%                                                  %%%
$$
\tilde{\Lambda}_8=\{\lambda\, \vert \, A_1=-A_2 \,,\, B_1=A_3 \,,\, B_2=-B_3\}\,.
$$
%                                                  %%%
So
%                                                  %%%
$$
 \left|  \int_{\tilde{\Lambda}_1}  [A_1(\vec{a}, \lambda) B_1(\vec{b}, \lambda) -  A_2(\vec{a}, \lambda) B_2(\vec{c}, \lambda)] \rho({\lambda}) d\lambda \right| 
$$
%                                                  %%%
$$    
\hspace{30pt} \leq \int_{\tilde{\Lambda}_1}  [1 +  A_3(\vec{b}, \lambda) B_3(\vec{c}, \lambda)] \rho({\lambda}) d\lambda
$$
%                                                  %%%
$$   
\hspace{30pt}=Z(\tilde{\Lambda}_1) - Z(\tilde{\Lambda}_1)\cos\theta_{ab}\cos\theta_{ac}\,,
$$
%                                                  %%%
where $Z(\tilde{\Lambda}_1)$ is the measure of the set $\tilde{\Lambda}_1$, and the integral of the product $A_3 B_3 \rho({\lambda})$ results in $-Z(\tilde{\Lambda}_1)\cos\theta_{ab}\cos\theta_{ac}$ given that, when $\lambda$ belongs to $\tilde{\Lambda}_1$, the functions $A_3$ and $B_3$ are correlated precisely by the constraints used to build the tables \ref{t1} - \ref{t6}.

Since we are not as familiar with the constraints in $\tilde{\Lambda}_2$ as those in $\tilde{\Lambda}_1$ we will perform the next integral in more detail.

%                                                  %%%
$$
 \left| \int_{\tilde{\Lambda}_2}  [A_1(\vec{a}, \lambda) B_1(\vec{b}, \lambda) -  A_2(\vec{a}, \lambda) B_2(\vec{c}, \lambda)] \rho({\lambda}) d\lambda \right|
$$ 
%                                                  %%%
$$
\hspace{107pt}  =  \left|  \int_{\tilde{\Lambda}_2}  A_1(\vec{a}, \lambda) B_1(\vec{b}, \lambda) [ 1 +  A_3(\vec{b}, \lambda) B_2(\vec{c}, \lambda)] \rho({\lambda}) d\lambda \right|,
$$
%                                                  %%%
given the first two constraints, which turns to
%                                                  %%%
$$   
\hspace{30pt} \leq  \int_{\tilde{\Lambda}_2}  [ 1 -  A_3(\vec{b}, \lambda) B_3(\vec{c}, \lambda)] \rho({\lambda}) d\lambda\,,
$$
%                                                  %%%
by use of the last constraint. We can determine this integral by using the correlations given in Table \ref{t7},
%                                                  %%%            table 1

\begin{table}[H]
\setlength\extrarowheight{10pt}
\caption{Joint probabilities of $A_3(\vec{b}, \lambda)$ and $B_3(\vec{c}, \lambda)$, when $\lambda$ belongs to $\tilde{\Lambda}_2$.} 
\label{t7}
\begin{center}
\begin{tabular}{|c|c|c|c|c|}
\hline
\diag{0.1em}{55pt}{\small{$A_2(\vec{a},\lambda)$}}{\small{\color{carmin}{$-A_3(\vec{b},\lambda)$}}}&1&-1 \\ [10pt]
\hline
1 &  \cen{0.1em}{55pt}{$\hspace{3pt}\frac{1}{2}$}{\color{carmin}{$\sin^2\frac{\theta_{ab}}{2}\hspace{5pt}$}} &  \cen{0.1em}{55pt}{$\hspace{3pt}\frac{1}{2}$}{\color{carmin}{$\cos^2\frac{\theta_{ab}}{2}\hspace{5pt}$}} \\ [10pt]
\hline
-1&\cen{0.1em}{55pt}{$\hspace{3pt}\frac{1}{2}$}{\color{carmin}{$\cos^2\frac{\theta_{ab}}{2}\hspace{5pt}$}} & \cen{0.1em}{55pt}{$\hspace{3pt}\frac{1}{2}$}{\color{carmin}{$\sin^2\frac{\theta_{ab}}{2}\hspace{5pt}$}} \\ [10pt]
\hline
\end{tabular}
\end{center}

\begin{center}
\begin{tabular}{|c|c|c|c|c|}
\hline
\diag{0.2em}{55pt}{\small{$A_2(\vec{a},\lambda)$}}{\small{\color{carmin}{$-B_3(\vec{c},\lambda)$}}}&1&-1 \\ 
\hline
1 &  \cen{0.1em}{55pt}{$\hspace{3pt}\frac{1}{2}$}{\color{carmin}{$\sin^2\frac{\theta_{ac}}{2}\hspace{5pt}$}} &  \cen{0.1em}{55pt}{$\hspace{3pt}\frac{1}{2}$}{\color{carmin}{$\cos^2\frac{\theta_{ac}}{2}\hspace{5pt}$}} \\ [10pt]
\hline
-1&\cen{0.1em}{55pt}{$\hspace{3pt}\frac{1}{2}$}{\color{carmin}{$\cos^2\frac{\theta_{ac}}{2}\hspace{5pt}$}} & \cen{0.1em}{55pt}{$\hspace{3pt}\frac{1}{2}$}{\color{carmin}{$\sin^2\frac{\theta_{ac}}{2}\hspace{5pt}$}} \\ [10pt]
\hline
\end{tabular}
\end{center}
\end{table}

%                                                  %%%
\noindent obtaining, once again, 
%                                                  %%%
$$   
\int_{\tilde{\Lambda}_2}  [ 1 -  A_3(\vec{b}, \lambda) B_3(\vec{c}, \lambda)] \rho({\lambda}) d\lambda =Z(\tilde{\Lambda}_2) - Z(\tilde{\Lambda}_2)\cos\theta_{ab}\cos\theta_{ac}\,.
$$
%                                                  %%%

Following the same procedure, one can verify that
%                                                  %%%
$$
\hspace{-20pt} \left|  \int_{\tilde{\Lambda}_i}  [A_1(\vec{a}, \lambda) B_1(\vec{b}, \lambda) -  A_2(\vec{a}, \lambda) B_2(\vec{c}, \lambda)] \rho({\lambda}) d\lambda \right|
$$
%                                                  %%%
$$   
\hspace{100pt}\leq Z(\tilde{\Lambda}_i) - Z(\tilde{\Lambda}_i)\cos\theta_{ab}\cos\theta_{ac}
$$
%                                                  %%%
$\forall i$. Adding all these integrals over $i$, and normalising to the volume of $\Lambda$, i.e. 
%                                                  %%%
$$ 
\sum_{i=1}^{8} Z(\tilde{\Lambda}_i) = 1\,,
$$
%                                                  %%%
yields the value $1 - \cos\theta_{ab}\cos\theta_{ac}$.

This shows that the inequality the two expectation values must satisfy, when assuming no specific relation between functions $A$ and $B$, is:
%                                                  %%%
\begin{equation}
\left| E(\vec{a}, \vec{b})-E(\vec{a}, \vec{c})\right| \leq 1 - \cos\theta_{ab}\cos\theta_{ac}\,,
\label{Bl}
\end{equation}
%                                                  %%%
where $\cos\theta_{ab}\cos\theta_{ac}$ is just a quantity, not an expectation value of a specific scenario.

We have already shown that Quantum Mechanics' predictions and experimental results always  satisfy inequality (\ref{Bl}). 

\section{Conclusions}
Local realism can be recovered for Quantum Mechanics when the factuality assumption is taken into account. We conclude this by showing that: 

The usual application of Bell's inequality to experiments is not a proof of the non-local nature of reality, in that in a factual universe Bell's inequality cannot be derived for the conditions of the built experiments. 

There is an inequality that can be experimentally tested, that is the Bell-like inequality we constructed in section \ref{Blike}, (eq. (\ref{Bl})). This inequality is always satisfied by Quantum Mechanics' predictions, and thus by the known experimental results.

\begin{figure}[h]
	\begin{center}
		\scalebox{0.06}{\includegraphics{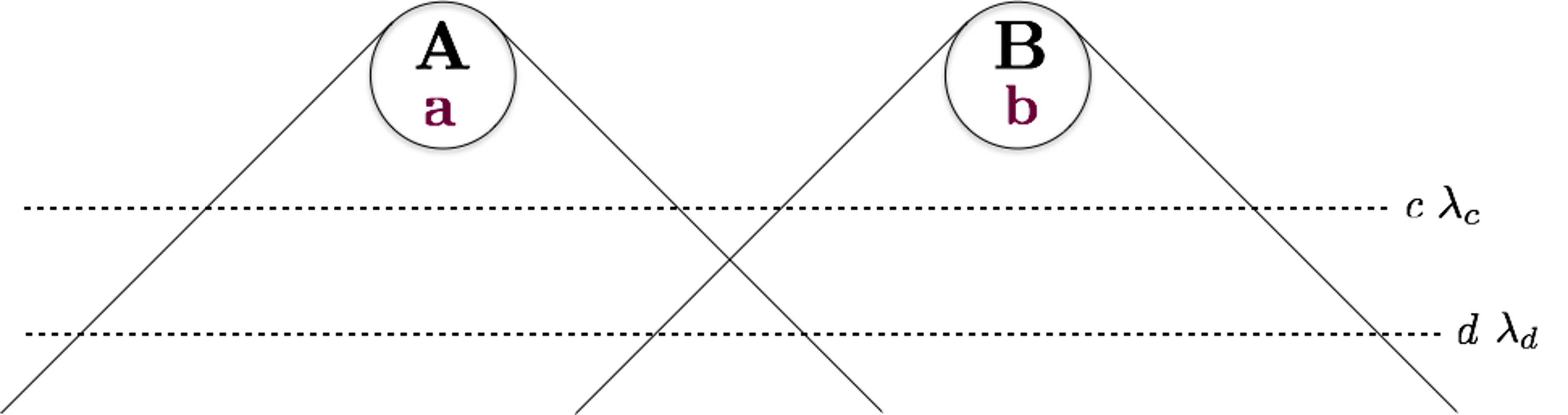}}
	\end{center}
	\caption{Events {\bf A} and {\bf B}, space-like separated, with results $A$ and $B$ from detectors setup at conditions (angles) $\vec{a}$ and $\vec{b}$, together with their past light-cones. The set $(c,\,\lambda_c)$, with $c$ the values of any variables that describe the experimental setup, and $\lambda_c$ the values of any additional (hidden) variables necessary to obtain a complete theory, completely specifies the results as it totally screens each result from the other's past light-cone. Figure adapted from figure 6 of J.S. Bell: ``La Nouvelle Cuisine''~\cite{BellNC}.}
	\label{LightCones}
\end{figure}

Our factuality assumption implies a common cause on the detectors and particle creation process, which is encoded in the hidden variables. This exploits the so-called {\it freedom of choice loophole}, which appears when questioning independence of the detector settings, from the hidden variables that emerge at the creation of the entangled states\cite{Larsson}. Let $A$, $B$, $\in\, \{-1,\,1\}$ denote the values of the detectors' results, and $\vec{a}$, $\vec{b}$, the angles at which detectors are set. Let $c$ stand for values of any variables that describe the experimental setup, and $\lambda_c$ for values of any additional (hidden) variables necessary to obtain a complete theory. We define the pair $(d,\,\lambda_d)$ in the same manner as $(c,\,\lambda_c)$, but at an earlier time (see Fig.~\ref{LightCones}).

% FIGURE 1
	
With these definitions, the joint probability $\mathcal{P}(A,\,B)$ of obtaining the particular results $A,\ B$ from the detectors is an explicit function of $\vec{a},\,\vec{b},\,c,\,\lambda_c$, which we denote by
$$
\mathcal{P}(A,\,B) = \mathcal{P}(A,\,B;\ \vec{a},\,\vec{b},\,c,\,\lambda_c)\,.
$$

In order to derive Bell's inequality, it is {\it necessary} to assume
$$
\mathcal{P}(A,\,B;\ \vec{a},\,\vec{b},\,c,\,\lambda_c) = \mathcal{P}(A;\ \vec{a},\,c,\,\lambda_c)\, \mathcal{P}(B;\ \vec{b},\,c,\,\lambda_c)\,,
$$
which is justified by invoking locality and the fact that $(A,\,\vec{a})$ and $(B,\,\vec{b})$ are space-like separated, i.e., $(A,\,\vec{a})$ does not depend on $(B,\,\vec{b})$, nor $(B,\,\vec{b})$ on $(A,\,\vec{a})$, but only on their own local conditions and on their causal past, which is contained in $(c,\,\lambda_c)$.

An {\it additional} assumption in the derivation is that $\vec{a}$ and $\vec{b}$ may be chosen freely, by the freedom-of-choice assumption.

But in the factuality scheme, a truly deterministic scheme, $(c,\,\lambda_c)$ contains information about $\vec{a}$ and about $\vec{b}$, {\it and about the correlation between $\vec{a}$ and $\vec{b}$} through $(d,\,\lambda_d)$, which itself contains information from the intersection of the past light-cones of $\mathbf{A}$ and $\mathbf{B}$. That is,
$$
c = c(d), \ \text{and}\ \lambda_c = \lambda_c(\lambda_d)
$$
so that, in fact, $\vec{a} = \vec{a}(\vec{b})$ and $\vec{b} = \vec{b}(\vec{a})$ through $c$ and $\lambda_c$.

Note that the factuality scheme does not discard the chaos or pseudo-randomness exhibited by complex systems: chaos is totally deterministic in essence, and it is only our inability to measure with absolute precision what prevents us from predicting the system's behaviour at all times. As for pseudo-randomness, it is algorithmic and therefore deterministic.

In this scheme the results $A$ and $B$ are predetermined by the hidden variables $\lambda_d$ at some point in their (near or far) past, independently of the spatial separation of the subsystems, because backward light-cones necessarily intersect. It is worth mentioning that the original experiments of Aspect~\cite{Aspect, Aspect1} and the much improved experiment of Weihs~\cite{Weihs} to address locality, do not resolve the freedom-of-choice loophole under the factuality scenario. Nor, for that matter, do the more recent experiments of Zeilinger's groups~\cite{Zeilinger1, Zeilinger2, Giustina} with ever-increasing space-like separations between subsystems, under the scenario here presented. The reader may also wish to see~\cite{Shalm}.

It seems, then, that a local interpretation of QM may be built. More work is in order, particularly on the mechanism of entanglement within this scenario. This is under current consideration.

\appendix
\section{} %Apéndice A
\label{ap}

We start from:
%                                                  %%%            
$$
\mathcal{F} (\lambda, t_1) = (\vec{o}_{A}(\lambda, t_1),\vec{o}_{B}(\lambda, t_1)) = (\pm\vec{a},\pm\vec{b})\,,
$$
%                                                  %%%
and we define:
%                                                  %%%            
$$
A_1(\vec{a},\lambda) \equiv \rm{sign}(\vec{o}_{A}(\lambda, t_1))
$$
%                                                  %%%              
and 
%                                                  %%%            
$$
B_1(\vec{b},\lambda) \equiv \rm{sign}(\vec{o}_{B}(\lambda, t_1))\,,
$$
%                                                  %%%
where we carried the subscript $1$ to distinguish these functions from the ones defined by  $\mathcal{F} (\lambda, t_2)$. Now:
%                                                  %%%            
$$
\mathcal{F} (\lambda, t_2) = (\vec{o}_{A}(\lambda, t_2),\vec{o}_{B}(\lambda, t_2)) = (\pm\vec{a},\pm\vec{c})\,,
$$
%                                                  %%%
so we can simultaneously define:
%                                                  %%%            
$$
A_2(\vec{a},\lambda) \equiv \rm{sign}(\vec{o}_{A}(\lambda, t_2))
$$
%                                                  %%%              
and 
%                                                  %%%            
$$
B_2(\vec{c},\lambda) \equiv \rm{sign}(\vec{o}_{B}(\lambda, t_2))\,.
$$
%                                                  %%%
And finally: 
%                                                  %%%            
$$
A_3(\vec{b},\lambda) \equiv \rm{sign}(\vec{o}_{A}(\lambda, t_3))
$$
%                                                  %%%              
and 
%                                                  %%%            
$$
B_3(\vec{c},\lambda) \equiv \rm{sign}(\vec{o}_{B}(\lambda, t_3))\,.
$$
%                                                  %%%
Now, of course functions $A_i$ and $B_i$ defined this way are not necessarily identical to those defined by the first path, just because $\vec{o}_{A}(\lambda, t_3)$ is not necessarily the same as $\vec{o}_{A_3}(\lambda, t_1)$, etc. The thing is that, once one defines a set of functions $\{A_1, B_1, A_2, B_2, A_3, B_3\}$, function $A_1(\vec{a},\lambda)$ \emph{can} be different from $A_2(\vec{a},\lambda)$ (and so forth) and this is the argument we use in the rest of our development. 

\section{} %Apéndice B
\label{ap2}
Bell parts from equation (\ref{1}),
%                                                  %%%
$$
\left| E(\vec{a}, \vec{b})-E(\vec{a}, \vec{c})\right|  = 
\left|  \int_{\Lambda}  [A_1(\vec{a}, \lambda) B_1(\vec{b}, \lambda) -  A_2(\vec{a}, \lambda) B_2(\vec{c}, \lambda)] \rho({\lambda}) d\lambda \right|
$$
%                                                  %%%
and makes his first assumption,
%                                                  %%%
$$
A_1(\vec{a}, \lambda) = A_2(\vec{a}, \lambda)\,;
$$
%                                                  %%%
then equation (\ref{1}) turns to:
%                                                  %%%
$$
\left| E(\vec{a}, \vec{b})-E(\vec{a}, \vec{c})\right|  = \left|\int_\Lambda  A_1(\vec{a}, \lambda) B_1(\vec{b}, \lambda) \left[1 - B_1(\vec{b}, \lambda) B_2(\vec{c}, \lambda)\right] \rho(\lambda)d\lambda \right|,
$$
%                                                  %%%
where he uses the fact that $B_1 B_1 =1$. Now, taking the absolute value function into the integral and using the fact that $\left|A_1 B_1\right|=1$ his last equation turns to:
%                                                  %%%
\begin{equation}
\left| E(\vec{a}, \vec{b})-E(\vec{a}, \vec{c})\right| \leq \int_\Lambda \left| \left[ 1 - B_1(\vec{b}, \lambda)  B_2(\vec{c}, \lambda) \right]  \rho(\lambda)\right| d\lambda\,,
\label{3} 
\end{equation}
%                                                  %%%
but what is inside the absolute value function is always positive, so he just discards the bars. Next comes his second assumption,
%                                                  %%%
$$
B_1(\vec{b}, \lambda)=-A_3(\vec{b}, \lambda)\,,
$$
%                                                  %%%
so equation (\ref{3}) becomes:
%                                                  %%%
\begin{equation}
\left| E(\vec{a}, \vec{b})-E(\vec{a}, \vec{c})\right| \leq \int_\Lambda  \left[ 1 + A_3(\vec{b}, \lambda)  B_2(\vec{c}, \lambda) \right]  \rho(\lambda) d\lambda\,.
\label{4} 
\end{equation}
%                                                  %%%
And finally, he takes a third assumption,
%                                                  %%%
$$
B_2(\vec{c}, \lambda)=B_3(\vec{c}, \lambda)\,,
$$
%                                                  %%%
then equation (\ref{4}) turns to:
%                                                  %%%
$$
\left| E(\vec{a}, \vec{b})-E(\vec{a}, \vec{c})\right|  {\leq} \int_\Lambda \left[ 1 + A_3(\vec{b}, \lambda)    B_3(\vec{c}, \lambda) \right] \rho(\lambda)d\lambda\,,
$$
%                                                  %%%
which takes him to his final step,
%                                                  %%%
$$
\int_\Lambda \left[ 1 + A_3(\vec{b}, \lambda)    B_3(\vec{c}, \lambda) \right] \rho(\lambda)d\lambda = 1 + E(\vec{b}, \vec{c})\,,
$$
%                                                  %%%
concluding,
%                                                  %%%
$$
\left| E(\vec{a}, \vec{b})-E(\vec{a}, \vec{c})\right| \leq 1 + E(\vec{b}, \vec{c})\,.
$$
%                                                  %%%

\section{} %Apéndice C
\label{ap3}

We begin by building a table of probabilities for the first scenario (detector settings $\vec{a}$ and $\vec{b}$), under the following knowledge: the probability of getting either $+1$ or $-1$ when measuring the spin projection of particle $A$ is $\frac{1}{2}$, but once one of those is guaranteed, say $+1$, the probability of getting $+1$  when measuring the spin projection of particle $B$ is $\sin^2\frac{\theta_{ab}}{2}$ and the probability of getting $-1$ is $\cos^2\frac{\theta_{ab}}{2}$. So we have the joint probabilities shown in Table \ref{t1}.  
%                                                  %%%            table C1
\begin{table}[H]
\setlength\extrarowheight{10pt}
\caption{Joint probabilities for experiment 1.}
\label{t1}
\begin{center}
\begin{tabular}{|c|c|c|}
\hline
\diag{0.1em}{55pt}{\small{$A_1(\vec{a},\lambda)$}}{\small{$B_1(\vec{b},\lambda)$}}&1&-1 \\ [10pt]
\hline
1 & \cen{0.1em}{55pt}{$\hspace{3pt}\frac{1}{2}$}{$\sin^2\frac{\theta_{ab}}{2}\hspace{5pt}$} & \cen{0.1em}{55pt}{$\hspace{3pt}\frac{1}{2}$}{$\cos^2\frac{\theta_{ab}}{2}\hspace{5pt}$}\\ [10pt]
\hline
-1&\cen{0.1em}{55pt}{$\hspace{3pt}\frac{1}{2}$}{$\cos^2\frac{\theta_{ab}}{2}\hspace{5pt}$} & \cen{0.1em}{55pt}{$\hspace{3pt}\frac{1}{2}$}{$\sin^2\frac{\theta_{ab}}{2}\hspace{5pt}$}  \\ [10pt]
\hline
\end{tabular}
\end{center}
\end{table}
%                                                  %%%

Now, the assumption $A_1(\vec{a}, \lambda) = A_2(\vec{a}, \lambda)$ invites us to substitute $A_1$ for $A_2$ and the assumption $B_1(\vec{b}, \lambda) = -A_3(\vec{b}, \lambda)$ allows us to substitute $B_1$ for $-A_3$ , turning Table \ref{t1} into Table \ref{t3}.
%                                                  %%%            table C2
\begin{table}[H]
\setlength\extrarowheight{10pt}
\caption{Joint probabilities  under the assumptions $A_1(\vec{a}, \lambda)=A_2(\vec{a}, \lambda)$ and $B_1(\vec{b},\lambda)=-A_3(\vec{b}, \lambda)$.}
\label{t3}
\begin{center}
\begin{tabular}{|c|c|c|c|c|}
\hline
\diag{0.1em}{55pt}{\small{$A_2(\vec{a},\lambda)$}}{\small{\color{carmin}{$-A_3(\vec{b},\lambda)$}}}&1&-1 \\ [10pt]
\hline
1 &  \cen{0.1em}{55pt}{$\hspace{3pt}\frac{1}{2}$}{\color{carmin}{$\sin^2\frac{\theta_{ab}}{2}\hspace{5pt}$}} &  \cen{0.1em}{55pt}{$\hspace{3pt}\frac{1}{2}$}{\color{carmin}{$\cos^2\frac{\theta_{ab}}{2}\hspace{5pt}$}} \\ [10pt]
\hline
-1&\cen{0.1em}{55pt}{$\hspace{3pt}\frac{1}{2}$}{\color{carmin}{$\cos^2\frac{\theta_{ab}}{2}\hspace{5pt}$}} & \cen{0.1em}{55pt}{$\hspace{3pt}\frac{1}{2}$}{\color{carmin}{$\sin^2\frac{\theta_{ab}}{2}\hspace{5pt}$}} \\ [10pt]
\hline
\end{tabular}
\end{center}
\end{table}
%                                                  %%%

The joint probabilities for experiment 2 are built accordingly and result in the top of Table \ref{t4}. Taking into account the assumption $B_2(\vec{c}, \lambda) = B_3(\vec{c}, \lambda)$ one gets the bottom of Table \ref{t4}.
%                                                  %%%            table C3
\begin{table}[H]
\setlength\extrarowheight{10pt}
\caption{Top: joint probabilities for experiment 2. Bottom: same, under the assumption $B_2(\vec{c}, \lambda) = B_3(\vec{c}, \lambda)$.}
\label{t4}
\begin{center}
\begin{tabular}{|c|c|c|c|c|}
\hline
\diag{0.2em}{55pt}{\small{$A_2(\vec{a},\lambda)$}}{\small{$B_2(\vec{c},\lambda)$}}&1&-1 \\ 
\hline
1 & \cen{0.1em}{55pt}{$\hspace{3pt}\frac{1}{2}$}{$\sin^2\frac{\theta_{ac}}{2}\hspace{5pt}$} & \cen{0.1em}{55pt}{$\hspace{3pt}\frac{1}{2}$} {$\cos^2\frac{\theta_{ac}}{2}\hspace{5pt}$} \\ 
\hline
-1&\cen{0.1em}{55pt}{$\hspace{3pt}\frac{1}{2}$}{$\cos^2\frac{\theta_{ac}}{2}\hspace{5pt}$} &\cen{0.1em}{55pt}{$\hspace{3pt}\frac{1}{2}$} {$\sin^2\frac{\theta_{ac}}{2}\hspace{5pt}$} \\ 
\hline
\end{tabular}
\end{center}

\begin{center}
\begin{tabular}{|c|c|c|c|c|}
\hline
\diag{0.2em}{55pt}{\small{$A_2(\vec{a},\lambda)$}}{\small{\color{carmin}{$B_3(\vec{c},\lambda)$}}}&1&-1 \\ 
\hline
1 &  \cen{0.1em}{55pt}{$\hspace{3pt}\frac{1}{2}$}{\color{carmin}{$\sin^2\frac{\theta_{ac}}{2}\hspace{5pt}$}} &  \cen{0.1em}{55pt}{$\hspace{3pt}\frac{1}{2}$}{\color{carmin}{$\cos^2\frac{\theta_{ac}}{2}\hspace{5pt}$}} \\ [10pt]
\hline
-1&\cen{0.1em}{55pt}{$\hspace{3pt}\frac{1}{2}$}{\color{carmin}{$\cos^2\frac{\theta_{ac}}{2}\hspace{5pt}$}} & \cen{0.1em}{55pt}{$\hspace{3pt}\frac{1}{2}$}{\color{carmin}{$\sin^2\frac{\theta_{ac}}{2}\hspace{5pt}$}} \\ [10pt]
\hline
\end{tabular}
\end{center}
\end{table}
%                                                  %%%

Table \ref{t6} just brings together Table \ref{t3} and the bottom of Table \ref{t4}. We will use this to compute the joint probabilities of $A_3(\vec{b}, \lambda)$ and $B_3(\vec{c}, \lambda)$.
%                                                  %%%            table C4
\begin{table}[H]
\setlength\extrarowheight{10pt}
\caption{Joint probabilities of $A_3(\vec{b}, \lambda)$ and $B_3(\vec{c}, \lambda)$.} 
\label{t6}
\begin{center}
\begin{tabular}{|c|c|c|c|c|}
\hline
\diag{0.1em}{55pt}{\small{$A_2(\vec{a},\lambda)$}}{\small{\color{carmin}{$-A_3(\vec{b},\lambda)$}}}&1&-1 \\ [10pt]
\hline
1 &  \cen{0.1em}{55pt}{$\hspace{3pt}\frac{1}{2}$}{\color{carmin}{$\sin^2\frac{\theta_{ab}}{2}\hspace{5pt}$}} &  \cen{0.1em}{55pt}{$\hspace{3pt}\frac{1}{2}$}{\color{carmin}{$\cos^2\frac{\theta_{ab}}{2}\hspace{5pt}$}} \\ [10pt]
\hline
-1&\cen{0.1em}{55pt}{$\hspace{3pt}\frac{1}{2}$}{\color{carmin}{$\cos^2\frac{\theta_{ab}}{2}\hspace{5pt}$}} & \cen{0.1em}{55pt}{$\hspace{3pt}\frac{1}{2}$}{\color{carmin}{$\sin^2\frac{\theta_{ab}}{2}\hspace{5pt}$}} \\ [10pt]
\hline
\end{tabular}
\end{center}

\begin{center}
\begin{tabular}{|c|c|c|c|c|}
\hline
\diag{0.2em}{55pt}{\small{$A_2(\vec{a},\lambda)$}}{\small{\color{carmin}{$B_3(\vec{c},\lambda)$}}}&1&-1 \\ 
\hline
1 &  \cen{0.1em}{55pt}{$\hspace{3pt}\frac{1}{2}$}{\color{carmin}{$\sin^2\frac{\theta_{ac}}{2}\hspace{5pt}$}} &  \cen{0.1em}{55pt}{$\hspace{3pt}\frac{1}{2}$}{\color{carmin}{$\cos^2\frac{\theta_{ac}}{2}\hspace{5pt}$}} \\ [10pt]
\hline
-1&\cen{0.1em}{55pt}{$\hspace{3pt}\frac{1}{2}$}{\color{carmin}{$\cos^2\frac{\theta_{ac}}{2}\hspace{5pt}$}} & \cen{0.1em}{55pt}{$\hspace{3pt}\frac{1}{2}$}{\color{carmin}{$\sin^2\frac{\theta_{ac}}{2}\hspace{5pt}$}} \\ [10pt]
\hline
\end{tabular}
\end{center}
\end{table}
%                                                  %%%
\noindent
\vspace{7pt}
\noindent The procedure is as follows:
\vspace{8pt}

\noindent $A_2(\vec{a}, \lambda)=+1$ for $\lambda$ in a certain set, say $\Lambda_+$, and from Table \ref{t6} if $\lambda \in \Lambda_+$, then the probability that $A_3(\vec{b}, \lambda)=1$ is $\cos^2({\theta_{ab}}/{2})$ and the probability that $A_3(\vec{b}, \lambda)=-1$ is $\sin^2({\theta_{ab}}/{2})$, while the probability that $B_3(\vec{b}, \lambda)=1$ is $\sin^2({\theta_{ac}}/{2})$ and the probability that $B_3(\vec{b}, \lambda)=-1$ is $\cos^2({\theta_{ac}}/{2})$. So, for $\lambda \in \Lambda_+$  the probability of getting the same sign in both functions $A_3$ and $B_3$ is:

\vspace{-2pt}
%                                                  %%%           
$$
	 \mathcal{P}(A_3\cdot B_3 =1) = \cos^2\frac{\theta_{ab}}{2}\sin^2\frac{\theta_{ac}}{2} + \sin^2\frac{\theta_{ab}}{2}\cos^2\frac{\theta_{ac}}{2}\,,
$$
%                                                  %%%

\vspace{3pt}
\noindent and the probability of getting opposite signs is: 

\vspace{-2pt}
%                                                  %%%           
$$
	 \mathcal{P}(A_3\cdot B_3 =-1) = \cos^2\frac{\theta_{ab}}{2}\cos^2\frac{\theta_{ac}}{2} + \sin^2\frac{\theta_{ab}}{2}\sin^2\frac{\theta_{ac}}{2}\,. 
$$
%                                                  %%%

\noindent Similarly, if $\lambda \in \Lambda_-$,

\vspace{-2pt}
%                                                  %%%           
$$
	 \mathcal{P}(A_3\cdot B_3 = 1) = \sin^2\frac{\theta_{ab}}{2}\cos^2\frac{\theta_{ac}}{2} + \cos^2\frac{\theta_{ab}}{2}\sin^2\frac{\theta_{ac}}{2} 
$$
%                                                  %%%

\vspace{3pt}
\noindent and

\vspace{-2pt}
%                                                  %%%           
$$
	 \mathcal{P}(A_3\cdot B_3 =-1) = \sin^2\frac{\theta_{ab}}{2}\sin^2\frac{\theta_{ac}}{2} + \cos^2\frac{\theta_{ab}}{2}\cos^2\frac{\theta_{ac}}{2}\,.
$$
%                                                  %%% 

\vspace{3pt}
But at the same time, the probability that $\lambda \in \Lambda_+$ is ${1}/{2}$ as is the probability that $\lambda \in \Lambda_-$, so we must multiply all the four last equations by $1/2$ and then add them to obtain:  

\vspace{-2pt}
%                                                  %%%           
\begin{equation}
	 \mathcal{P}(A_3\cdot B_3 = 1) = \cos^2\frac{\theta_{ab}}{2}\sin^2\frac{\theta_{ac}}{2} + \sin^2\frac{\theta_{ab}}{2}\cos^2\frac{\theta_{ac}}{2}
	 \label{pp}
\end{equation}
%                                                  %%%

\vspace{3pt}
\noindent and

\vspace{-2pt}
%                                                  %%%           
\begin{equation}
	 \mathcal{P}(A_3\cdot B_3 =-1) = \sin^2\frac{\theta_{ab}}{2}\sin^2\frac{\theta_{ac}}{2} + \cos^2\frac{\theta_{ab}}{2}\cos^2\frac{\theta_{ac}}{2}\,.
	 	 \label{pm}
\end{equation}
%                                                  %%% 

\vspace{3pt}
Functions with the probability distributions given by equations (\ref{pp}) and (\ref{pm}) describe an experiment in which the expectation value of the correlation between these two functions would be:

\vspace{-4pt}
%                                                  %%%           
$$
E(\vec{b},\vec{c}) =  \mathcal{P}(A_3\cdot B_3 =1)- \mathcal{P}(A_3\cdot B_3 = -1)
$$
%                                                  %%%
$$   
\hspace{37pt} = \cos^2\frac{\theta_{ab}}{2}\sin^2\frac{\theta_{ac}}{2} + \sin^2\frac{\theta_{ab}}{2}\cos^2\frac{\theta_{ac}}{2} 
$$
%                                                  %%%
$$   
\hspace{57.5pt} - \sin^2\frac{\theta_{ab}}{2}\sin^2\frac{\theta_{ac}}{2} - \cos^2\frac{\theta_{ab}}{2}\cos^2\frac{\theta_{ac}}{2}
$$
%                                                  %%%
$$   
\hspace{58pt} = (\cos^2\frac{\theta_{ab}}{2}-\sin^2\frac{\theta_{ab}}{2})(\sin^2\frac{\theta_{ac}}{2}-\cos^2\frac{\theta_{ac}}{2}) 
$$
%                                                  %%%
$$   
\hspace{-35pt} = -\cos\theta_{ab}\cos\theta_{ac}\,.
$$
%                                                  %%%

\section{} %Apéndice D
\label{ap4}

The inequality to be analysed is: 

\vspace{-5pt}
%                                                  %%%
$$
\vert -\cos\theta_{ab} + \cos\theta_{ac} \vert \leq 1 -\cos\theta_{ab}\cos\theta_{ac}\,,
$$
%                                                  %%%

\vspace{2pt}
\noindent which turns to

\vspace{-5pt}
%                                                  %%%
$$
\cos\theta_{ab}\cos\theta_{ac} -1 \leq  -\cos\theta_{ab} + \cos\theta_{ac} \leq 1 -\cos\theta_{ab}\cos\theta_{ac}\,.
$$
%                                                  %%%

\vspace{2pt}
\noindent The inequality on the left is satisfied iff

\vspace{-5pt}
%                                                  %%% 
$$
\cos \theta_{ab}\cos \theta_{ac} +  \cos \theta_{ab} \leq \cos \theta_{ac}   +1\,, 
$$
%                                                  %%%

\noindent or, equivalently, 

\vspace{-5pt}
%                                                  %%%
\begin{equation}
\cos \theta_{ab}(\cos \theta_{ac} +1) \leq \cos \theta_{ac}   +1\,, 
\label{ineq1}
\end{equation}
%                                                  %%%

\vspace{2pt}
\noindent and the inequality on the right is satisfied iff

\vspace{-5pt}
%                                                  %%%
$$
\cos \theta_{ac}  + \cos \theta_{ab}\cos \theta_{ac} \leq 1 + \cos \theta_{ab}\,, 
$$
%                                                  %%%

\vspace{2pt}
\noindent or, equivalently, 

\vspace{-5pt}
%                                                  %%%
\begin{equation} 
\cos \theta_{ac}(1 + \cos \theta_{ab})  \leq 1 + \cos \theta_{ab}\,.
\label{ineq2}
\end{equation} 
%                                                  %%%

\vspace{2pt}
\noindent Finally, equations (\ref{ineq1}) and (\ref{ineq2}) are both true iff

\vspace{-5pt}
%                                                  %%%
$$
	 \cos \theta_{ab} \leq 1 \hspace{7pt} \text{and} \hspace{5pt} \cos\theta_{ac} \leq 1\,,
$$
%                                                  %%%

\vspace{2pt}
\noindent which always holds. So inequality (\ref{eq}) is always satisfied.

%\section*{References}

\end{document}